\pdfoutput=1
\documentclass[prb,twocolumn,showpacs,aps]{revtex4}
\usepackage{graphicx,graphics,color,epsfig}
\usepackage{bm}
\usepackage{amsmath}
\usepackage{amssymb}
\usepackage{epstopdf}

\makeatletter

\newcommand{\Rmnum}[1]{\expandafter\@slowromancap\romannumeral #1@}
\makeatother

\usepackage{color}

\begin{document}

\title{Possible superconductivity in Sr$_{2}$IrO$_{4}$ probed by quasiparticle interference}

\author{Yi Gao,$^{1}$ Tao Zhou,$^{2}$ Huaixiang Huang,$^{3}$ and Qiang-Hua Wang $^{4}$}
\affiliation{$^{1}$Department of Physics and Institute of Theoretical Physics,
Nanjing Normal University, Nanjing, 210023, China\\
$^{2}$College of Science, Nanjing University of Aeronautics and
Astronautics, Nanjing, 210016, China\\
$^{3}$Department of Physics, Shanghai University, Shanghai, 200444, China\\
$^{4}$National Laboratory of Solid State Microstructures, Nanjing University, Nanjing, 210093, China}

\begin{abstract}
Based on the possible superconducting (SC) pairing symmetries recently proposed, the quasiparticle interference (QPI) patterns in electron- and hole-doped Sr$_{2}$IrO$_{4}$ are theoretically investigated. In the electron-doped case, the QPI spectra can be explained based on a model similar to the octet model of the cuprates while in the hole-doped case, both the Fermi surface topology and the sign of the SC order parameter resemble those of the iron pnictides and there exists a QPI vector resulting from the interpocket scattering between the electron and hole pockets. In both cases, the evolution of the QPI vectors with energy and their behaviors in the nonmagnetic and magnetic impurity scattering cases can well be explained based on the evolution of the constant-energy contours and the sign structure of the SC order parameter. The QPI spectra presented in this paper can be compared with future scanning tunneling microscopy experiments to test whether there are SC phases in electron- and hole-doped Sr$_{2}$IrO$_{4}$ and what the pairing symmetry is.
\end{abstract}


\maketitle

\emph{Introduction}.---Recently, the $5d$ transition metal oxide Sr$_{2}$IrO$_{4}$ has attracted much attention.\cite{randall,crawford,cao,kim,kim2,moon,watanabe2,ishii,wang,jkim,fujiyama,cetin,ye} In this material, the energy bands close to the Fermi level are mainly contributed by the $t_{2g}$ orbitals of Ir and it is in the $(t_{2g})^{5}$ configuration. On the one hand, due to the extended nature of $5d$ orbitals, Coulomb interaction $U$ for $5d$ electrons ($1-3$eV) is expected to be smaller than that for $3d$ electrons ($5-7$eV).\cite{watanabe2} On the other hand, the spin-orbit coupling (SOC) is considerably larger by a factor of $10$ in $5d$ than in $3d$.\cite{watanabe2} In this case, the strong SOC splits the $t_{2g}$ orbitals into an upper $J=\frac{1}{2}$ band and lower $J=\frac{3}{2}$ bands. In the parent compound, the $J=\frac{3}{2}$ bands are fully occupied while the $J=\frac{1}{2}$ band is half-filled. Meanwhile, the bandwidth of this $J=\frac{1}{2}$ band is much smaller than the original one in the absence of the SOC. Therefore, even a small $U$ can lead the system into a Mott insulator, making Sr$_{2}$IrO$_{4}$ an analog to the parent compound of the cuprates. This $J=\frac{1}{2}$ Mott insulating state is supported by several experiments.\cite{kim,kim2,jkim,ishii,fujiyama,cetin,moon,ye} The question is, whether doping Sr$_{2}$IrO$_{4}$ can induce superconductivity in analogy to the cuprates?

To resolve this issue, Refs. \onlinecite{watanabe} and \onlinecite{yyang} theoretically investigated the superconducting (SC) properties in both electron- and hole-doped Sr$_{2}$IrO$_{4}$. They found that, in the electron-doped case, a SC phase indeed exists and the pairing contains both intraorbital and interorbital components as well as both singlet and triplet components of $t_{2g}$ electrons, while the pairing symmetry on the Fermi surface is $d_{x^{2}-y^{2}}$-wave (or $d_{x^{2}-y^{2}}^{*}$-wave as denoted by Ref. \onlinecite{yyang}) and the pairing function respects time-reversal symmetry (TRS). On the other hand, in the hole-doped case, the Fermi surface topology changes and resembles that of the iron pnictides, with a electron pocket around the $\Gamma$ point and a hole pocket around the $M$ point. In this case, Ref. \onlinecite{watanabe} found that there is no SC phase while Ref. \onlinecite{yyang} concluded that a SC phase can also exist while the pairing function still respects TRS and the pairing symmetry is $s_{\pm}^{*}$-wave, similar to that of the iron pnictides.\cite{mazin}

In this paper, in order to search for an experimental test of the above two theories, we propose to measure the quasipariticle interference (QPI) patterns in both electron- and hole-doped Sr$_{2}$IrO$_{4}$ by scanning tunneling microscopy (STM). As we know, the QPI patterns are strongly influenced by the shape and evolution of the constant-energy contour (CEC), as well as the relative sign of the SC order parameter of the states connected by the QPI wave vectors.\cite{qpi,hoffman,hanaguri1,hanaguri2,mcelroy} Therefore, by measuring the QPI patterns, we can not only determine whether the SC phase exists in the electron- and hole-doped cases, but also the SC pairing symmetry.

\emph{Method}.---We start with the lattice model adopted in Refs.~\onlinecite{watanabe} and \onlinecite{yyang}, which takes the three $t_{2g}$ orbitals ($d_{xz}$, $d_{yz}$ and $d_{xy}$) of Ir into account. The Hamiltonian can be written as
\begin{eqnarray}
\label{h}
H&=&\frac{1}{2}\sum_{\mathbf{k}}\psi_{\mathbf{k}}^{\dag}M_{\mathbf{k}}\psi_{\mathbf{k}},\nonumber\\
\psi_{\mathbf{k}}^{\dag}&=&(c_{\mathbf{k}1\uparrow}^{\dag},c_{\mathbf{k}2\uparrow}^{\dag},c_{\mathbf{k}3\uparrow}^{\dag},c_{\mathbf{k}1\downarrow}^{\dag},c_{\mathbf{k}2\downarrow}^{\dag},c_{\mathbf{k}3\downarrow}^{\dag},\nonumber\\
& &c_{-\mathbf{k}1\uparrow},c_{-\mathbf{k}2\uparrow},c_{-\mathbf{k}3\uparrow},c_{-\mathbf{k}1\downarrow},c_{-\mathbf{k}2\downarrow},c_{-\mathbf{k}3\downarrow}),\nonumber\\
M_{\mathbf{k}}&=&\begin{pmatrix}
A_{\mathbf{k}}&\Delta_{0}D_{\mathbf{k}}\\\Delta_{0}D_{\mathbf{k}}^{\dag}&-A_{-\mathbf{k}}^{T}
\end{pmatrix},
\end{eqnarray}
where
\begin{eqnarray}
\label{Dk}
A_{\mathbf{k}}&=&\begin{pmatrix}
\epsilon_{1\mathbf{k}}&-\frac{i\lambda}{2}&0&0&0&\frac{i\lambda}{2}\\
\frac{i\lambda}{2}&\epsilon_{2\mathbf{k}}&0&0&0&-\frac{\lambda}{2}\\
0&0&\epsilon_{3\mathbf{k}}&-\frac{i\lambda}{2}&\frac{\lambda}{2}&0\\
0&0&\frac{i\lambda}{2}&\epsilon_{1\mathbf{k}}&\frac{i\lambda}{2}&0\\
0&0&\frac{\lambda}{2}&-\frac{i\lambda}{2}&\epsilon_{2\mathbf{k}}&0\\
-\frac{i\lambda}{2}&-\frac{\lambda}{2}&0&0&0&\epsilon_{3\mathbf{k}}
\end{pmatrix},\nonumber\\
D_{\mathbf{k}}&=&\begin{pmatrix}
0&0&\gamma_{\mathbf{k}}^{1}&g_{\mathbf{k}}^{1}&\gamma_{\mathbf{k}}^{2}&0\\
0&0&\gamma_{\mathbf{k}}^{3}&-\gamma_{\mathbf{k}}^{2}&g_{\mathbf{k}}^{2}&0\\
-\gamma_{\mathbf{k}}^{1}&-\gamma_{\mathbf{k}}^{3}&0&0&0&g_{\mathbf{k}}^{3}\\
-g_{\mathbf{k}}^{1}&\gamma_{\mathbf{k}}^{2}&0&0&0&-\gamma_{\mathbf{k}}^{1}\\
-\gamma_{\mathbf{k}}^{2}&-g_{\mathbf{k}}^{2}&0&0&0&\gamma_{\mathbf{k}}^{3}\\
0&0&-g_{\mathbf{k}}^{3}&\gamma_{\mathbf{k}}^{1}&-\gamma_{\mathbf{k}}^{3}&0
\end{pmatrix},\nonumber\\
\epsilon_{1\mathbf{k}}&=&-2t_{4}\cos k_{x}-2t_{5}\cos k_{y}-\mu,\nonumber\\
\epsilon_{2\mathbf{k}}&=&-2t_{5}\cos k_{x}-2t_{4}\cos k_{y}-\mu,\nonumber\\
\epsilon_{3\mathbf{k}}&=&-2t_{1}(\cos k_{x}+\cos k_{y})-4t_{2}\cos k_{x}\cos k_{y}\nonumber\\
& &-2t_{3}(\cos 2k_{x}+\cos 2k_{y})+\mu_{xy}-\mu.
\end{eqnarray}
Here $c_{\mathbf{k}1\uparrow}^{\dag}$, $c_{\mathbf{k}2\uparrow}^{\dag}$ and $c_{\mathbf{k}3\uparrow}^{\dag}$ create a spin-up electron with momentum $\mathbf{k}$ in the $d_{xz}$, $d_{yz}$ and $d_{xy}$ orbitals, respectively. $A_{\mathbf{k}}$ stands for the tight-binding part of the Hamiltonian in the presence of the SOC, with $\lambda$ being the SOC strength. $(t_{1},t_{2},t_{3},t_{4},t_{5},\mu_{xy},\lambda)=(0.36,0.18,0.09,0.37,0.06,-0.36,0.5)$ and $\mu$ is the chemical potential which is adjusted according to the electron filling $n$. $D_{\mathbf{k}}$ describes the pairing term of the Hamiltonian whose explicit expression is given later and we set $\Delta_{0}=0.05$ (unless otherwise specified).

When a single impurity is located at the origin, the impurity Hamiltonian can be written as
\begin{eqnarray}
\label{himp}
H_{imp}&=&\sum_{l=1}^{3}\sum_{\sigma=\uparrow,\downarrow}(V_{s}+s_{\sigma}V_{m})c_{0l\sigma}^{\dag}c_{0l\sigma}\nonumber\\
&=&\frac{1}{N}\sum_{l=1}^{3}\sum_{\sigma=\uparrow,\downarrow}\sum_{\mathbf{k},\mathbf{k}^{'}}(V_{s}+s_{\sigma}V_{m})c_{\mathbf{k}l\sigma}^{\dag}c_{\mathbf{k}^{'}l\sigma},
\end{eqnarray}
with $N$ being the system size ($396\times396$ throughout the paper) and $s_{\sigma}=1(-1)$ for $\sigma=\uparrow(\downarrow)$. We consider both nonmagnetic and magnetic impurity scattering, diagonal in the orbital basis and with a scattering strength $V_{s}$ and $V_{m}$ for the nonmagnetic and magnetic cases, respectively. Following the standard $T$-matrix procedure,~\cite{zhu} the Green's function matrix is defined as
\begin{eqnarray}
\label{gt}
g(\mathbf{k},\mathbf{k}^{'},\tau)=-\langle T_{\tau}\psi_{\mathbf{k}}(\tau)\psi_{\mathbf{k}^{'}}^{\dag}(0)\rangle,
\end{eqnarray}
and
\begin{eqnarray}
\label{gw}
g(\mathbf{k},\mathbf{k}^{'},\omega)&=&\delta_{\mathbf{k}\mathbf{k}^{'}}g_{0}(\mathbf{k},\omega)\nonumber\\
&&+g_{0}(\mathbf{k},\omega)T(\omega)g_{0}(\mathbf{k}^{'},\omega).
\end{eqnarray}
Here $g_{0}(\mathbf{k},\omega)$ is the Green's function in the absence of the impurity and can be written as
\begin{eqnarray}
\label{g0}
g_{0}(\mathbf{k},\omega)&=&[(\omega+i0^{+})I-M_{\mathbf{k}}]^{-1},\nonumber\\
T(\omega)&=&[I-\frac{\Omega}{N}\sum_{\mathbf{q}}g_{0}(\mathbf{q},\omega)]^{-1}\frac{\Omega}{N},\nonumber\\
\end{eqnarray}
where $I$ is a $12\times12$ unit matrix and
\begin{eqnarray}
\label{u}
\Omega_{lm}=\begin{cases}
V_{s}+V_{m}&\text{$l=m=1,2,3$},\\
V_{s}-V_{m}&\text{$l=m=4,5,6$},\\
-(V_{s}+V_{m})&\text{$l=m=7,8,9$},\\
-(V_{s}-V_{m})&\text{$l=m=10,11,12$},\\
0&\text{otherwise}.
\end{cases}
\end{eqnarray}
The experimentally measured local density of states (LDOS) is expressed as
\begin{eqnarray}
\label{rour}
\rho(\mathbf{r},\omega)&=&-\frac{1}{\pi}\sum_{l=1}^{3}\sum_{\sigma=\uparrow,\downarrow}{\rm Im}\langle\langle c_{\mathbf{r}l\sigma}|c_{\mathbf{r}l\sigma}^{\dag}\rangle\rangle_{\omega+i0^{+}}\nonumber\\
&=&-\frac{1}{\pi N}\sum_{m=1}^{6}\sum_{\mathbf{k},\mathbf{k}^{'}}{\rm Im}\Big{[}g_{mm}(\mathbf{k},\mathbf{k}^{'},\omega)e^{-i(\mathbf{k}-\mathbf{k}^{'})\cdot\mathbf{r}}\Big{]},\nonumber\\
\end{eqnarray}
and its Fourier transform is defined as $\rho(\mathbf{q},\omega)=\sum_{\mathbf{r}}\rho(\mathbf{r},\omega)e^{i\mathbf{q}\cdot\mathbf{r}}$, which can be expressed as
\begin{widetext}
\begin{eqnarray}
\label{rouq}
\rho(\mathbf{q},\omega)&=&-\frac{1}{2\pi}\sum_{m=1}^{6}\sum_{\mathbf{k}}{\rm Im}[g_{mm}(\mathbf{k},\mathbf{k}+\mathbf{q},\omega)+g_{mm}(\mathbf{k},\mathbf{k}-\mathbf{q},\omega)]\nonumber\\
&&+i{\rm Re}[g_{mm}(\mathbf{k},\mathbf{k}+\mathbf{q},\omega)-g_{mm}(\mathbf{k},\mathbf{k}-\mathbf{q},\omega)].
\end{eqnarray}
\end{widetext}

Since the system is even under $\mathbf{k}\rightarrow -\mathbf{k}$ ($D_{\mathbf{k}}$ is also an even function of $\mathbf{k}$ as can be seen later), Eq. (\ref{rouq}) can be simplified as
\begin{eqnarray}
\label{rouqsimplified}
\rho(\mathbf{q},\omega)&=&-\frac{1}{\pi}\sum_{m=1}^{6}\sum_{\mathbf{k}}{\rm Im}g_{mm}(\mathbf{k},\mathbf{k}+\mathbf{q},\omega),
\end{eqnarray}
and the contribution from the spin up and spin down electrons can be expressed as
\begin{eqnarray}
\label{spinup&dw}
\rho_{\uparrow}(\mathbf{q},\omega)&=&-\frac{1}{\pi}\sum_{m=1}^{3}\sum_{\mathbf{k}}{\rm Im}g_{mm}(\mathbf{k},\mathbf{k}+\mathbf{q},\omega),\nonumber\\
\rho_{\downarrow}(\mathbf{q},\omega)&=&-\frac{1}{\pi}\sum_{m=4}^{6}\sum_{\mathbf{k}}{\rm Im}g_{mm}(\mathbf{k},\mathbf{k}+\mathbf{q},\omega).
\end{eqnarray}

Here we need to clarify what to measure in the STM experiment. If the impurity scattering is weak, then $T(\omega)\approx\frac{\Omega}{N}$. In this case, if $V_{s}\neq0$ and $V_{m}=0$, we have $\rho_{\uparrow}(\mathbf{q},\omega)=\rho_{\downarrow}(\mathbf{q},\omega)$ since the system respects TRS. On the other hand, if $V_{s}=0$ and $V_{m}\neq0$, TRS is broken and now for $\mathbf{q}\neq0$, we have $\rho_{\uparrow}(\mathbf{q},\omega)=-\rho_{\downarrow}(\mathbf{q},\omega)$, leading to $\rho(\mathbf{q},\omega)=0$. Therefore, in the STM experiment, people should measure the spin-resolved LDOS, either $\rho_{\uparrow}(\mathbf{r},\omega)$ or $\rho_{\downarrow}(\mathbf{r},\omega)$, to get a nontrivial QPI spectrum.

\begin{figure}
\includegraphics[width=1\linewidth]{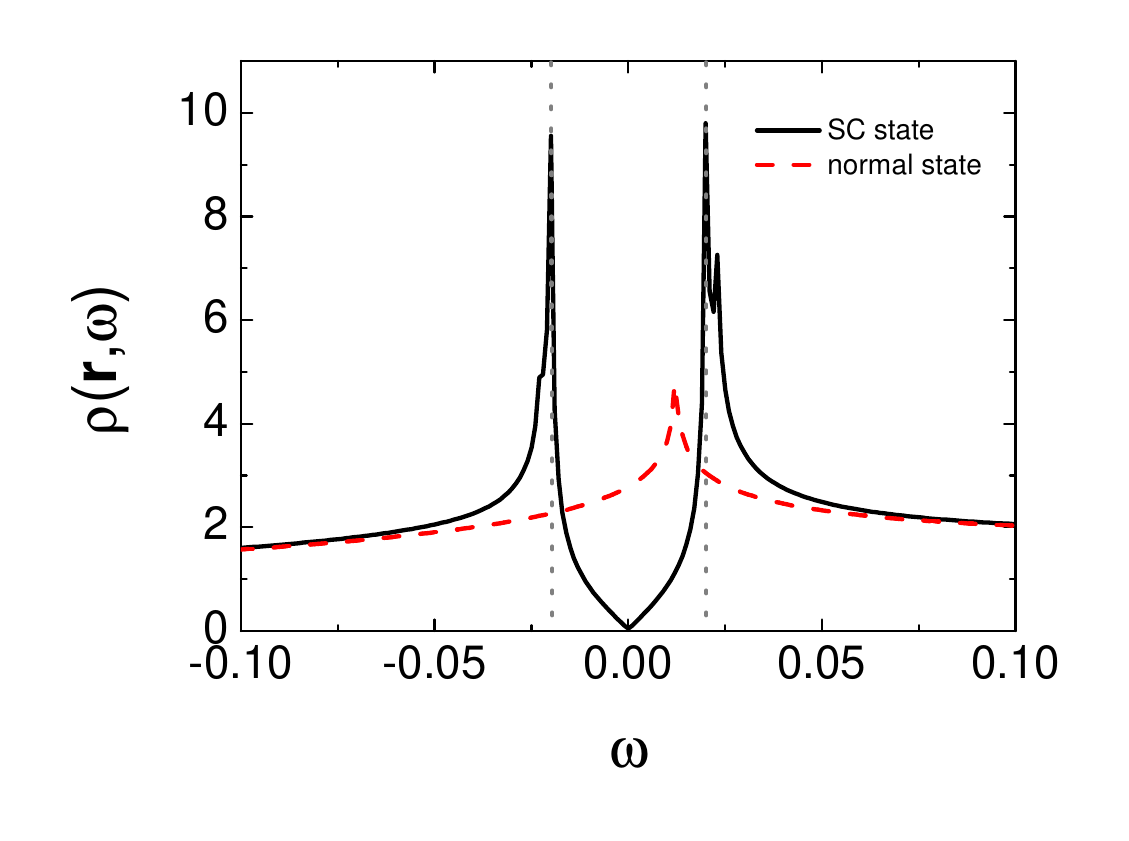}
 \caption{\label{dos_5.2} (Color online) At $n=5.2$, $\rho(\mathbf{r},\omega)$ as a function of $\omega$, in the absence of the impurity. The gray dotted lines denote the position of the two SC coherence peaks, located at $\pm\Delta$ ($\Delta\approx0.4\Delta_{0}$).}
\end{figure}

\begin{figure*}
\includegraphics[width=1\linewidth]{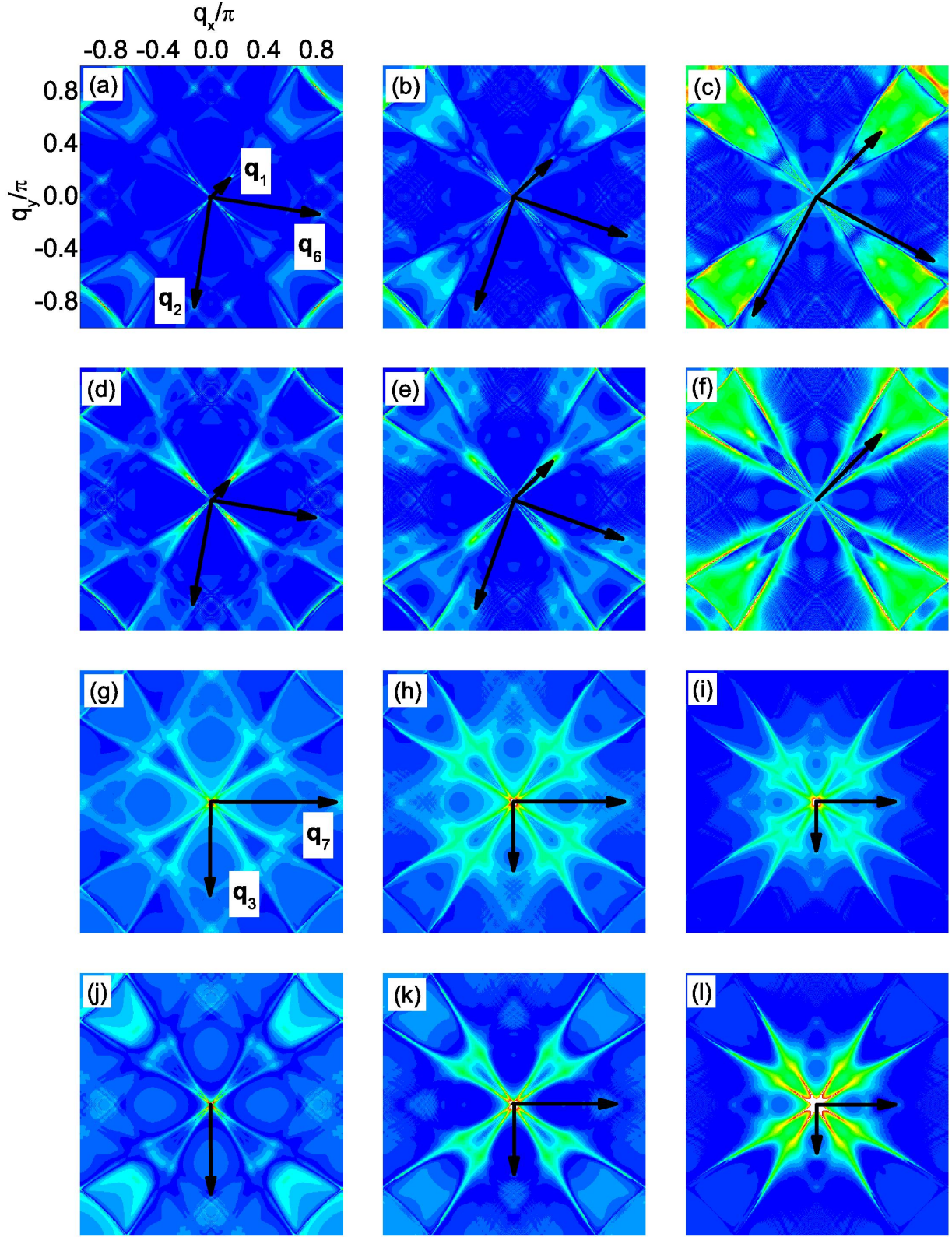}
 \caption{\label{qpi_5.2} (Color online) At $n=5.2$, $|\rho_{\uparrow}(\mathbf{q},\omega)|$ at fixed $\omega$. The point at $\mathbf{q}=0$ is neglected in order to show weaker features at other wave vectors. (a-f) $\omega/\Delta=-0.25,-0.5,-0.75,0.25,0.5,0.75$, for the nonmagnetic impurity scattering. (g-l) are the same as (a-f), but for the magnetic impurity scattering.}
\end{figure*}

\begin{figure}
\includegraphics[width=1\linewidth]{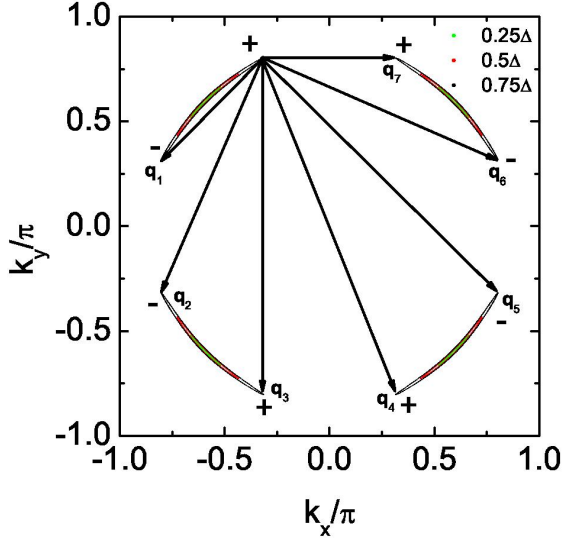}
 \caption{\label{cec_5.2} (Color online) At $n=5.2$, the CEC at $|\omega|/\Delta=0.25$ (green), $0.5$ (red) and $0.75$ (black). $\mathbf{q}_{1},\mathbf{q}_{2},\ldots,\mathbf{q}_{7}$ are characteristic QPI wave vectors connecting the tips of the CEC. The $+$ and $-$ denote the sign of the SC order parameter on the CEC.}
\end{figure}

\begin{figure}
\includegraphics[width=1\linewidth]{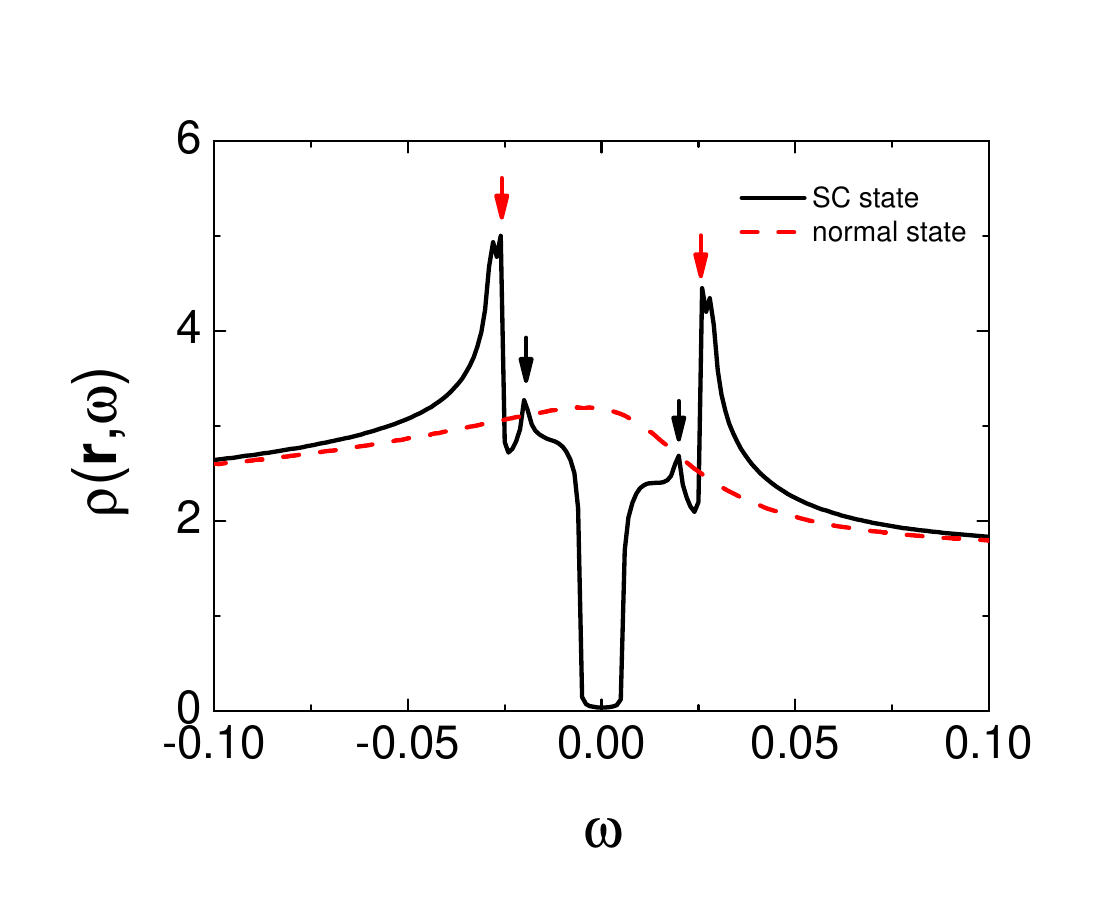}
 \caption{\label{dos_4.25} (Color online) The same as Fig. \ref{dos_5.2}, but at $n=4.25$.}
\end{figure}

\begin{figure}
\includegraphics[width=1\linewidth]{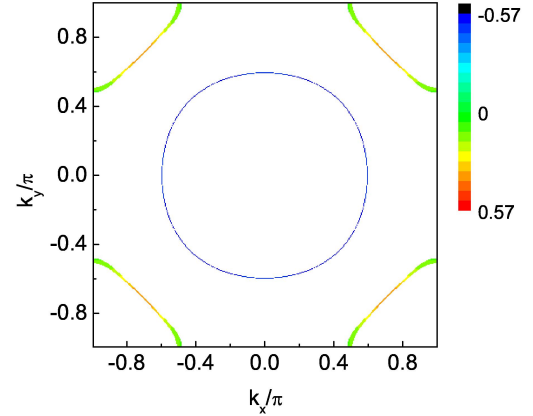}
 \caption{\label{fs_4.25} (Color online) At $n=4.25$, the pairing function $D_{\mathbf{k}}$ projected onto the Fermi surface.}
\end{figure}

\begin{figure*}
\includegraphics[width=1\linewidth]{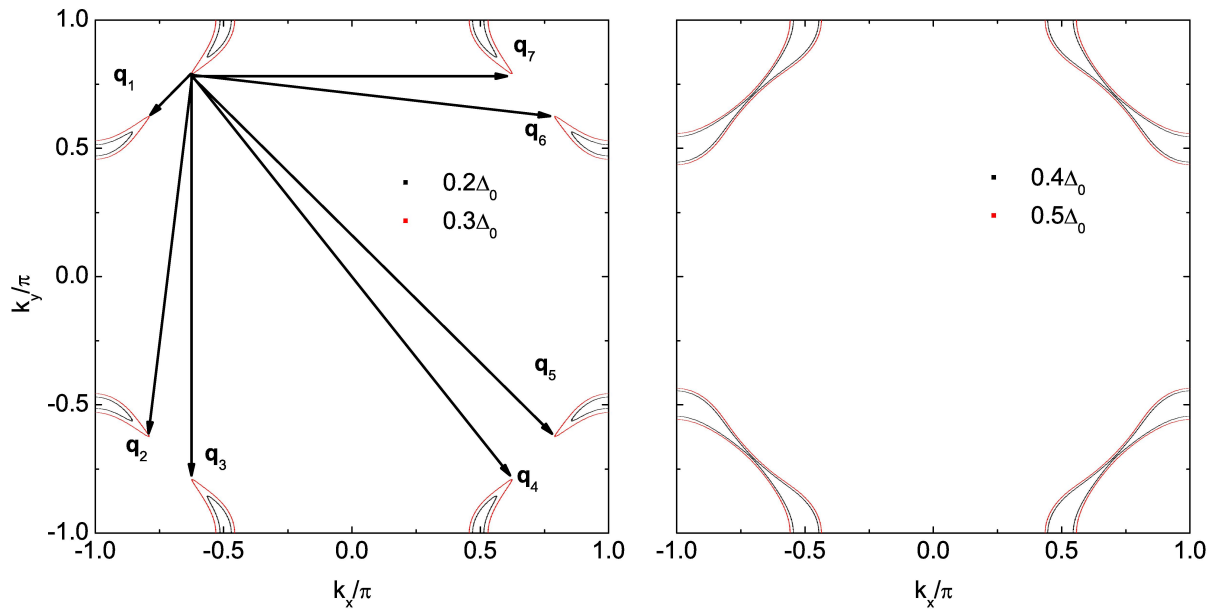}
 \caption{\label{cec_4.25_0.2_to_0.5} (Color online) The same as Fig. \ref{cec_5.2}, but at $n=4.25$. Left: $|\omega|=0.2\Delta_{0}$ (black) and $0.3\Delta_{0}$ (red). Right: $|\omega|=0.4\Delta_{0}$ (black) and $0.5\Delta_{0}$ (red).}
\end{figure*}

\begin{figure}
\includegraphics[width=1\linewidth]{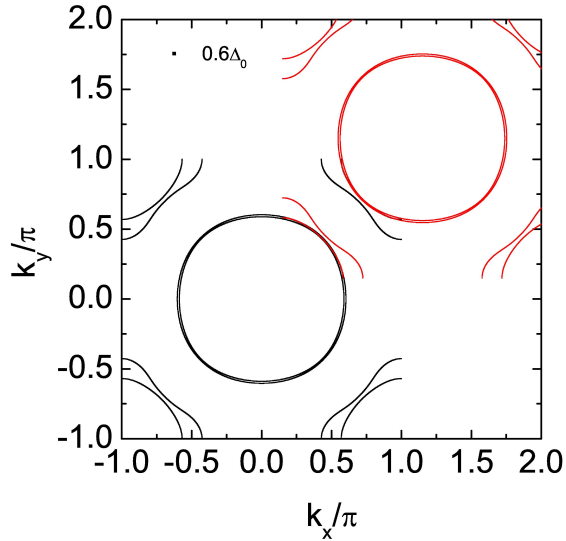}
 \caption{\label{cec_4.25_0.6} (Color online) The same as Fig. \ref{cec_4.25_0.2_to_0.5}, but at $|\omega|=0.6\Delta_{0}$. The red curves are displaced by $(1.15\pi,1.15\pi)$ from the black ones.}
\end{figure}

\begin{figure}
\includegraphics[width=1\linewidth]{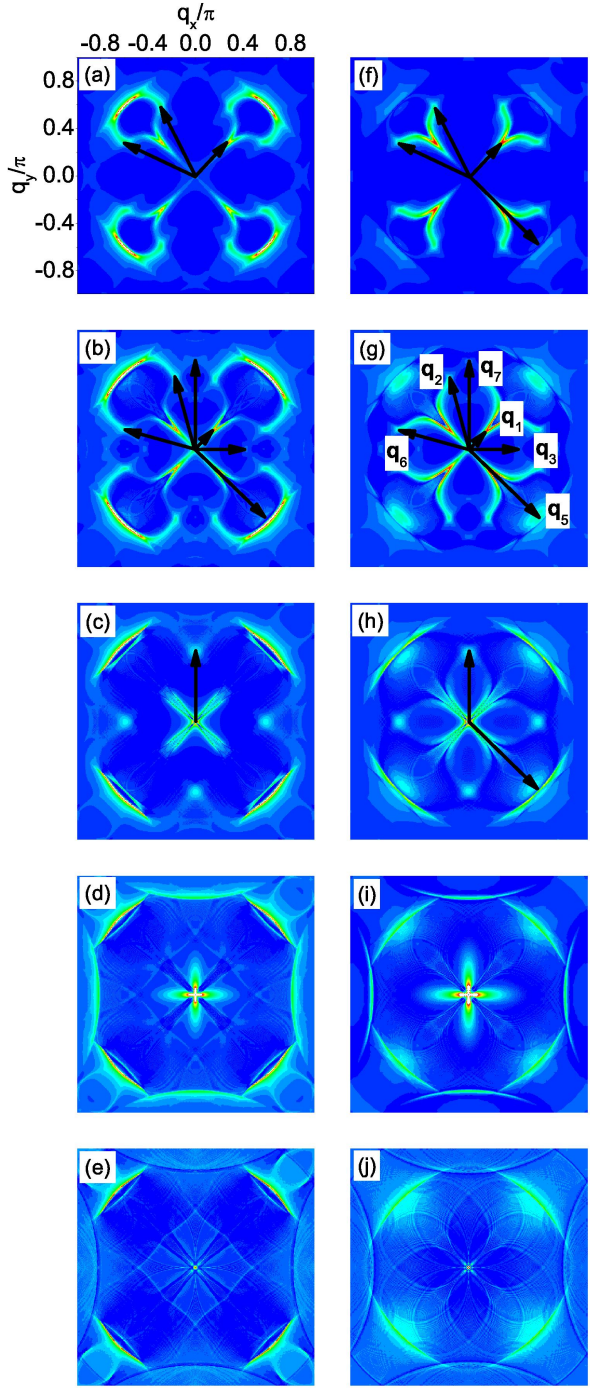}
 \caption{\label{qpi_4.25_mag} (Color online) At $n=4.25$ and for magnetic impurity scattering, $|\rho_{\uparrow}(\mathbf{q},\omega)|$ at fixed $\omega$. The point at $\mathbf{q}=0$ is neglected. (a-e): $|\omega|/\Delta_{0}=-0.2,-0.3,-0.4,-0.5,-0.6$. (f-j): $|\omega|/\Delta_{0}=0.2,0.3,0.4,0.5,0.6$.}
\end{figure}

\begin{figure}
\includegraphics[width=1\linewidth]{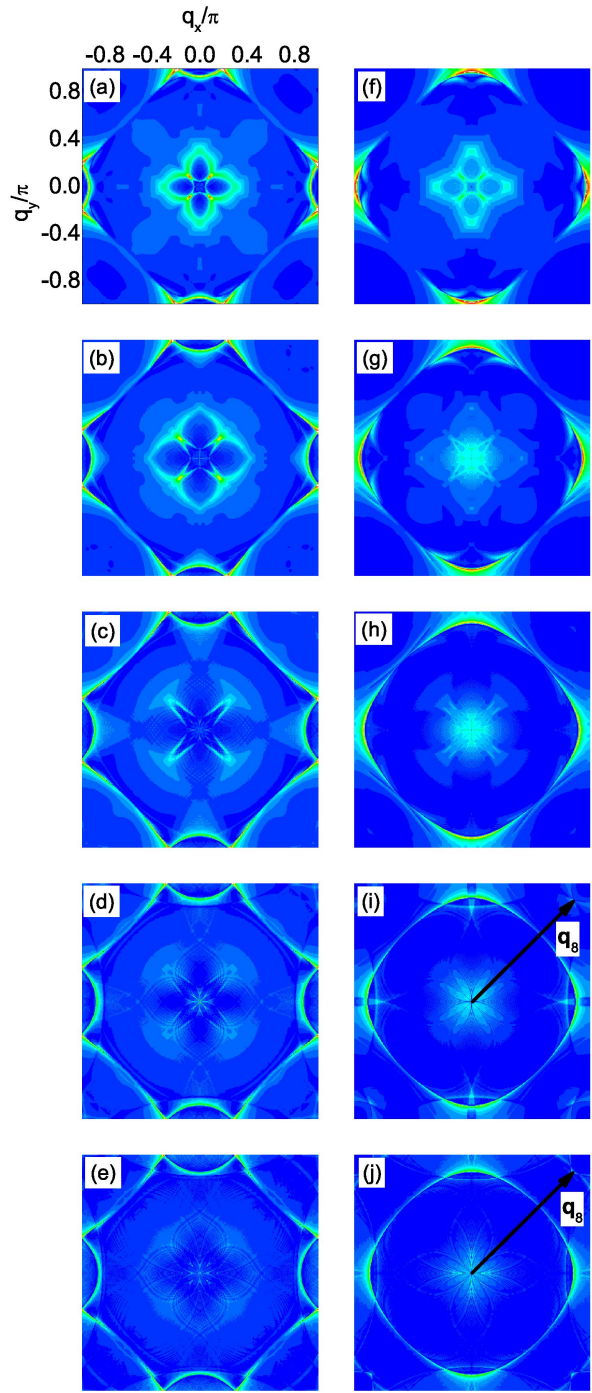}
 \caption{\label{qpi_4.25_nonmag} (Color online) The same as Fig. \ref{qpi_4.25_mag}, but for nonmagnetic impurity scattering.}
\end{figure}

\emph{Results}.---At $n=5.2$, the electron-doped case, the pairing functions $g_{\mathbf{k}}$ and $\gamma_{\mathbf{k}}$ in Eq. (\ref{Dk}) can be expressed as \cite{yyang}
\begin{eqnarray}
\label{Dk_5.2}
g_{\mathbf{k}}^{1}&=&-0.35+0.2\cos k_{y}-0.08\cos k_{x},\nonumber\\
g_{\mathbf{k}}^{2}&=&0.35-0.2\cos k_{x}+0.08\cos k_{y},\nonumber\\
g_{\mathbf{k}}^{3}&=&0.07(\cos k_{y}-\cos k_{x}),\nonumber\\
\gamma_{\mathbf{k}}^{1}&=&-i(0.15\cos k_{y}-0.12\cos k_{x}),\nonumber\\
\gamma_{\mathbf{k}}^{2}&=&0.23i(\cos k_{x}-\cos k_{y}),\nonumber\\
\gamma_{\mathbf{k}}^{3}&=&-(0.15\cos k_{x}-0.12\cos k_{y}).
\end{eqnarray}
The LDOS in the absence of the impurity is homogeneous in real space and is shown in Fig. \ref{dos_5.2}. Two SC coherence peaks are located at $\pm\Delta$ where $\Delta\approx0.4\Delta_{0}$ and the spectrum is V-shaped in the vicinity of $\omega=0$, indicating the nodal gap structure, consistent with the $d_{x^{2}-y^{2}}^{*}$-wave pairing symmetry.

In the presence of the impurity, we plot $|\rho_{\uparrow}(\mathbf{q},\omega)|$ in Fig. \ref{qpi_5.2} and several QPI wave vectors can be identified. For nonmagnetic impurity scattering [From Figs. \ref{qpi_5.2}(a) to \ref{qpi_5.2}(f)], three QPI wave vectors $\mathbf{q}_{1}$, $\mathbf{q}_{2}$ and $\mathbf{q}_{6}$ can be clearly seen evolving with energy.
$\mathbf{q}_{1}$ is located along the $(\pm1,\pm1)$ directions and moves away from the origin as $|\omega|$ increases. $\mathbf{q}_{2}$ and $\mathbf{q}_{6}$ are not located along the high-symmetry directions and they overlap after a $90$ degree rotation. Furthermore, they are not so obvious at $\omega/\Delta=0.75$ since they are masked by the high-intensity spots around them. In contrast, for magnetic impurity scattering [From Figs. \ref{qpi_5.2}(g) to \ref{qpi_5.2}(l)], $\mathbf{q}_{1}$, $\mathbf{q}_{2}$ and $\mathbf{q}_{6}$ become less clear and instead, another two vectors $\mathbf{q}_{3}$ and $\mathbf{q}_{7}$ can be identified evolving with energy. They are both located along the $(0,\pm1)$ and $(\pm1,0)$ directions and move towards the origin as $|\omega|$ increases.

The appearance and evolution of the above five QPI wave vectors can be understood from the evolution of the CEC. As we can see from Fig. \ref{cec_5.2}, the CEC of the electron-doped Sr$_{2}$IrO$_{4}$ is similar to the octet model of the cuprates \cite{qpi,hoffman,hanaguri1,zhu,mcelroy} and the expected QPI vectors should be those connecting the tips of the CEC, i.e., $\mathbf{q}_{1},\mathbf{q}_{2},\ldots,\mathbf{q}_{7}$ in this case. For example, at $|\omega|/\Delta=0.5$, $\mathbf{q}_{1},\mathbf{q}_{2},\ldots,\mathbf{q}_{7}$ shown in Fig. \ref{cec_5.2} are located at $(-0.295,-0.295)$, $(-0.295,0.839)$, $(0,0.544)$, $(0.866,0.544)$, $(-0.839,0.839)$, $(-0.839,-0.295)$ and $(0.866,0)$, agree quite well with those shown in Figs. \ref{qpi_5.2}(b), \ref{qpi_5.2}(e), \ref{qpi_5.2}(h) and \ref{qpi_5.2}(k), except that $\mathbf{q}_{4}$ and $\mathbf{q}_{5}$ cannot be identified. At $|\omega|/\Delta=0.25$ and $0.75$, the locations of the QPI vectors shown in Fig. \ref{cec_5.2} are also consistent with those in Fig. \ref{qpi_5.2} and for all the energies we investigated, $\mathbf{q}_{4}$ and $\mathbf{q}_{5}$ cannot be clearly seen, thus we neglect these two vectors in the following.

Next we discuss the implication of the QPI vectors on the sign of the SC order parameter. As we know, due to the effect of the SC coherence factors, those scattering between the states with the opposite (same) sign of the SC order parameters will be enhanced (suppressed) by nonmagnetic impurity. For magnetic impurity scattering, the situation is reversed. In electron-doped Sr$_{2}$IrO$_{4}$, since the pairing symmetry is assumed to be $d^{*}_{x^{2}-y^{2}}$-wave and the sign of the SC order parameter on the CEC is shown in Fig. \ref{cec_5.2} as $+$ and $-$. As we can see, $\mathbf{q}_{1}$, $\mathbf{q}_{2}$ and $\mathbf{q}_{6}$ are sign-reversing scattering processes while $\mathbf{q}_{3}$ and $\mathbf{q}_{7}$ are sign-preserving ones. Therefore, $\mathbf{q}_{1}$, $\mathbf{q}_{2}$ and $\mathbf{q}_{6}$ should be more discernable in the nonmagnetic impurity scattering case while $\mathbf{q}_{3}$ and $\mathbf{q}_{7}$ should be more distinct in the magnetic impurity scattering case. This is exactly what we obtain here as can be seen from Fig. \ref{qpi_5.2}. Therefore, the evolution of the QPI vectors with energy together with their different behaviors in the nonmagnetic and magnetic impurity scattering cases can help to determine whether the pairing symmetry is $d^{*}_{x^{2}-y^{2}}$-wave in electron-doped Sr$_{2}$IrO$_{4}$.

Here we need to point out that, Ref. \onlinecite{watanabe} assumed that the SC pairing is a pseudospin singlet formed by the $J=\frac{1}{2}$ Kramers doublet and the pairing symmetry is $d_{x^{2}-y^{2}}$-wave. In this case, the pairing term of the Hamiltonian can be written as $\Delta_{\mathbf{k}}a^{\dag}_{\mathbf{k}1\uparrow}a^{\dag}_{-\mathbf{k}1\downarrow}$, where $\Delta_{\mathbf{k}}=\frac{\Delta_{0}}{2}(\cos k_{x}-\cos k_{y})$ and $a^{\dag}_{\mathbf{k}1\uparrow}$ creates a pseudospin up electron with momentum $\mathbf{k}$ in the $J=\frac{1}{2}$ band. If we set $\Delta_{0}=0.02$ here, then the LDOS in the absence of the impurity is qualitatively the same as that shown in Fig. \ref{dos_5.2} and now we have $\Delta=\Delta_{0}$. In addition, the evolution of the CEC and the QPI spectra obtained are also similar to those in Figs. \ref{cec_5.2} and \ref{qpi_5.2}, respectively, indicating that the pairing functions adopted in Refs. \onlinecite{watanabe} and \onlinecite{yyang} share the same characteristics. As we can see in the limit of large SOC ($\lambda\rightarrow\infty$), $\Delta_{\mathbf{k}}a^{\dag}_{\mathbf{k}1\uparrow}a^{\dag}_{-\mathbf{k}1\downarrow}\propto\Delta_{\mathbf{k}}(c^{\dag}_{\mathbf{k}3\uparrow}+c^{\dag}_{\mathbf{k}2\downarrow}+ic^{\dag}_{\mathbf{k}1\downarrow})(c^{\dag}_{-\mathbf{k}3\downarrow}-c^{\dag}_{-\mathbf{k}2\uparrow}+ic^{\dag}_{-\mathbf{k}1\uparrow})$.
Although the pairing is a pseudospin singlet, it contains both intraorbital and interorbital components as well as both singlet and triplet components of $t_{2g}$ electrons and it respects the same symmetry as that shown in Eq. (\ref{Dk_5.2}). Therefore, for electron-doped Sr$_{2}$IrO$_{4}$, Refs. \onlinecite{watanabe} and \onlinecite{yyang} predicted similar SC phases.

Then we consider the hole-doped Sr$_{2}$IrO$_{4}$ at $n=4.25$. In this case, the pairing function proposed by Ref. \onlinecite{yyang} can be written as
\begin{eqnarray}
\label{Dk_4.25}
g_{\mathbf{k}}^{1}&=&-0.11-0.3(\cos k_{x}+\cos k_{y}),\nonumber\\
g_{\mathbf{k}}^{2}&=&g_{\mathbf{k}}^{1},\nonumber\\
g_{\mathbf{k}}^{3}&=&-0.21-0.14(\cos k_{x}+\cos k_{y}),\nonumber\\
\gamma_{\mathbf{k}}^{1}&=&i(0.17-0.02\cos k_{x}-0.12\cos k_{y}),\nonumber\\
\gamma_{\mathbf{k}}^{2}&=&i[0.19+0.04(\cos k_{x}+\cos k_{y})],\nonumber\\
\gamma_{\mathbf{k}}^{3}&=&-(0.17-0.12\cos k_{x}-0.02\cos k_{y}).
\end{eqnarray}
The LDOS in the absence of the impurity is shown in Fig. \ref{dos_4.25} and two pairs of SC coherence peaks are located at $\pm0.4\Delta_{0}$ and $\pm0.52\Delta_{0}$, as denoted by the black and red arrows, respectively, with a U-shaped profile close to $\omega=0$, indicating the full gap opening at this doping level. The pairing function $D_{\mathbf{k}}$ projected onto the Fermi surface is shown in Fig. \ref{fs_4.25}. As we can see, the pairing order parameter on the electron pocket around $\Gamma$ is negative and almost isotropic while on the hole pocket around $M$, it is positive and anisotropic. Therefore, the Fermi surface topology and the sign change of the pairing order parameter between the electron and hole pockets are very similar to the iron pnictides \cite{hanaguri2,mazin} and this pairing symmetry is dubbed as $s_{\pm}^{*}$-wave.

The evolution of the CEC with energy at this doping level is shown in Figs. \ref{cec_4.25_0.2_to_0.5} and \ref{cec_4.25_0.6}. As can be seen, at low energies ($|\omega|=0.2\Delta_{0}$ and $0.3\Delta_{0}$), the CEC exists around the $M$ point and the characteristic QPI vectors should be $\mathbf{q}_{1},\mathbf{q}_{2},\ldots,\mathbf{q}_{7}$ as shown in the left panel of Fig. \ref{cec_4.25_0.2_to_0.5}. $\mathbf{q}_{1}$ and $\mathbf{q}_{5}$ are located along the $(\pm1,\pm1)$ directions. $\mathbf{q}_{1}$ moves towards the origin as $|\omega|$ increases while $\mathbf{q}_{5}$ hardly evolves with energy. $\mathbf{q}_{2}$ and $\mathbf{q}_{6}$ are not located along the high-symmetry directions while $\mathbf{q}_{3}$ and $\mathbf{q}_{7}$ are both located along the $(0,\pm1)$ and $(\pm1,0)$ directions. In addition, $\mathbf{q}_{3}$ should move towards the origin with increasing $|\omega|$ while the situation for $\mathbf{q}_{7}$ is reversed. As $|\omega|$ increases to $0.4\Delta_{0}$, the tips of the two adjacent CECs touch each other. So in this case, $\mathbf{q}_{1}$ should disappear while $\mathbf{q}_{2}$, $\mathbf{q}_{3}$, $\mathbf{q}_{6}$ and $\mathbf{q}_{7}$ become equivalent. When $|\omega|\geq0.5\Delta_{0}$, the CEC evolves into closed contours where no tips exist, thus the above mentioned QPI vectors disappear. Here $\mathbf{q}_{1},\mathbf{q}_{2},\ldots,\mathbf{q}_{7}$ are all sign-preserving scattering processes, therefore they should be discernable only in the magnetic impurity scattering case. As $|\omega|$ increases to $0.6\Delta_{0}$, another CEC shows up around the $\Gamma$ point. In this case, a large portion of the CECs around the $M$ and $\Gamma$ points are quasinested with each other by a nesting vector $(\pm1.15\pi,\pm1.15\pi)$, as can be seen from Fig. \ref{cec_4.25_0.6}. In this case, there should exist a QPI vector located at around $(\pm0.85\pi,\pm0.85\pi)$ in the first Brillouin zone and since it is a sign-reversing scattering process, it should be observable only in the nonmagnetic impurity scattering case.\cite{hanaguri2}

To verify the above expectations, the QPI spectra are calculated and are plotted in Figs. \ref{qpi_4.25_mag} and \ref{qpi_4.25_nonmag}. For magnetic impurity scattering (see Fig. \ref{qpi_4.25_mag}), indeed we can identify the QPI vectors $\mathbf{q}_{1},\mathbf{q}_{2},\ldots,\mathbf{q}_{7}$, except that $\mathbf{q}_{4}$ cannot be clearly seen. The evolution of these vectors is also consistent with that derived from Fig. \ref{cec_4.25_0.2_to_0.5}, i.e., $\mathbf{q}_{1}$ locates along the $(\pm1,\pm1)$ directions and moves towards the origin with increasing $|\omega|$. $\mathbf{q}_{3}$ and $\mathbf{q}_{7}$ both locate along the $(0,\pm1)$ and $(\pm1,0)$ directions while they become equivalent with $\mathbf{q}_{2}$ and $\mathbf{q}_{6}$ at $|\omega|/\Delta_{0}=0.4$. Meanwhile, $\mathbf{q}_{5}$ barely evolves with energy and at $|\omega|/\Delta_{0}\geq0.5$, the above mentioned QPI vectors disappear. On the other hand, for nonmagnetic impurity scattering, as we can see from Fig. \ref{qpi_4.25_nonmag}, $\mathbf{q}_{1},\mathbf{q}_{2},\ldots,\mathbf{q}_{7}$ become less clear and instead, at $\omega/\Delta_{0}=0.5$ and $0.6$ [see Figs. \ref{qpi_4.25_nonmag}(i) and \ref{qpi_4.25_nonmag}(j)], another QPI vector $\mathbf{q}_{8}$ shows up at around $(\pm0.85\pi,\pm0.85\pi)$, which is resulted from the interpocket scattering between the electron and hole pockets as we mentioned above. Therefore, the locations of these QPI vectors and their behaviors in the nonmagnetic and magnetic impurity scattering cases are consistent with what we expected from the evolution of the CEC and the sign structure of the SC order parameter.

\emph{Summary}.---In summary, we have studied the QPI spectra in both electron- and hole-doped Sr$_{2}$IrO$_{4}$, by assuming the pairing symmetries proposed by Refs. \onlinecite{watanabe} and \onlinecite{yyang}. In the electron-doped case, we found that the pairing functions in Refs. \onlinecite{watanabe} and \onlinecite{yyang} are qualitatively the same and the QPI spectra can be explained based on a model similar to the octet model of the cuprates. On the other hand, for hole-doped Sr$_{2}$IrO$_{4}$, the QPI spectra in the SC phase resemble those of the iron pnictides where the interpocket scattering between the electron and hole pockets lead to a QPI vector locating at the nesting vector of these two pockets. In both cases, the evolution of the QPI vectors and their different behaviors in the nonmagnetic and magnetic impurity scattering cases can well be explained based on the evolution of the CEC and the sign structure of the SC order parameter. The QPI spectra presented in this paper can thus be compared with future STM experiments to test whether there are SC phases in electron- and hole-doped Sr$_{2}$IrO$_{4}$ and what the SC pairing symmetry is.

We thank Y. Xiong and Y. Yang for helpful discussions. This work was supported by NSFC (Grants No. 11204138, No. 11374005 and No. 11023002), the Ministry of Science and Technology of China (Grants No. 2011CBA00108 and 2011CB922101), NSF of Jiangsu Province of China (Grant No. BK2012450), NSF of Shanghai (Grant No. 13ZR1415400), SRFDP (Grant No. 20123207120005) and NCET (Grant No. NCET-12-0626). The numerical calculations in this paper have been done on the IBM Blade cluster system in the High Performance Computing Center (HPCC) of Nanjing University.


\begin{thebibliography}{99}

\bibitem{randall} J. J. Randall, L. Katz, and R. Ward, J. Am. Chem. Soc. \textbf{79}, 266 (1957).

\bibitem{crawford} M. K. Crawford, M. A. Subramanian, R. L. Harlow, J. A. Fernandez-Baca, Z. R. Wang, and D. C. Johnston, Phys. Rev. B \textbf{49}, 9198 (1994).

\bibitem{cao} G. Cao, J. Bolivar, S. McCall, J. E. Crow, and R. P. Guertin, Phys. Rev. B \textbf{57}, R11039 (1998).

\bibitem{kim} B. J. Kim, H. Jin, S. J. Moon, J.-Y. Kim, B.-G. Park, C. S. Leem, J. Yu, T. W. Noh, C. Kim, S.-J. Oh, J.-H. Park, V. Durairaj, G. Cao, and E. Rotenberg, Phys. Rev. Lett. \textbf{101}, 076402 (2008).

\bibitem{kim2} B. J. Kim, H. Ohsumi, T. Komesu, S. Sakai, T. Morita, H. Takagi, and T. Arima, Science \textbf{323}, 1329 (2009).

\bibitem{moon} S. J. Moon, H. Jin, W. S. Choi, J. S. Lee, S. S. A. Seo, J. Yu, G. Cao, T. W. Noh, and Y. S. Lee, Phys. Rev. B \textbf{80}, 195110 (2009).

\bibitem{watanabe2} H. Watanabe, T. Shirakawa, and S. Yunoki, Phys. Rev. Lett. \textbf{105}, 216401 (2010).

\bibitem{ishii} K. Ishii, I. Jarrige, M. Yoshida, K. Ikeuchi, J. Mizuki, K. Ohashi, T. Takayama, J. Matsuno, and H. Takagi, Phys. Rev. B \textbf{83}, 115121 (2011).

\bibitem{wang} F. Wang and T. Senthil, Phys. Rev. Lett. \textbf{106}, 136402 (2011).

\bibitem{jkim} J. Kim, D. Casa, M. H. Upton, T. Gog, Y.-J. Kim, J. F. Mitchell, M. van Veenendaal, M. Daghofer, J. van den Brink, G. Khaliullin, and B. J. Kim, Phys. Rev. Lett. \textbf{108}, 177003 (2012).

\bibitem{fujiyama} S. Fujiyama, H. Ohsumi, T. Komesu, J. Matsuno, B. J. Kim, M. Takata, T. Arima, and H. Takagi, Phys. Rev. Lett. \textbf{108}, 247212 (2012).

\bibitem{cetin} M. F. Cetin, P. Lemmens, V. Gnezdilov, D. Wulferding, D. Menzel, T. Takayama, K. Ohashi, and H. Takagi, Phys. Rev. B \textbf{85}, 195148 (2012).

\bibitem{ye} F. Ye, S. Chi, B. C. Chakoumakos, J. A. Fernandez-Baca, T. Qi, and G. Cao, Phys. Rev. B \textbf{87}, 140406 (2013).

\bibitem{watanabe} H. Watanabe, T. Shirakawa, and S. Yunoki, Phys. Rev. Lett. \textbf{110}, 027002 (2013).

\bibitem{yyang} Y. Yang, W.-S. Wang, J.-G. Liu, H. Chen, J.-H. Dai, and Q.-H. Wang, Phys. Rev. B \textbf{89}, 094518 (2014).

\bibitem{mazin} I. I. Mazin, D. J. Singh, M. D. Johannes, and M. H. Du, Phys. Rev. Lett. \textbf{101}, 057003 (2008).

\bibitem{qpi} Q. H. Wang and D. H. Lee, Phys. Rev. B {\bf 67}, 020511 (2003).

\bibitem{hoffman} J. E. Hoffman, K. McElroy, D.-H. Lee, K. M Lang, H. Eisaki, S. Uchida, and J. C. Davis, Science \textbf{297}, 1148 (2002).

\bibitem{mcelroy} K. McElroy, R. W. Simmonds, J. E. Hoffman, D.-H. Lee, J. Orenstein, H. Eisaki, S. Uchida, and J. C. Davis, Nature (London) \textbf{422}, 592 (2003).

\bibitem{hanaguri1} T. Hanaguri, Y. Kohsaka, M. Ono, M. Maltseva, P. Coleman, I. Yamada, M. Azuma, M. Takano, K. Ohishi, and H. Takagi, Science \textbf{323}, 923 (2009).

\bibitem{hanaguri2} T. Hanaguri, S. Niitaka, K. Kuroki, and H. Takagi, Science \textbf{328}, 474 (2010).

\bibitem{zhu} A. V. Balatsky, I. Vekhter, and J.-X. Zhu, Rev. Mod. Phys. \textbf{78}, 373 (2006).

\end{thebibliography}
\end{document}